\documentclass[aps,prb,twocolumn,groupedaddress,showpacs,floatfix]{revtex4-1}

\usepackage{subfigure}
\usepackage{graphicx}
\usepackage{bm}
\usepackage{amsmath}

\newcommand{\STO}{SrTiO$_3$}
\newcommand{\LAO}{LaAlO$_3$}

\newcommand{\etal}{\emph{et al.}}

\begin{document}

\title{Low-temperature dependence of the thermo-magnetic transport properties of the \STO/\LAO~ interface.}



\author{S. Lerer}
\email[]{shaharl3@post.tau.ac.il}
\affiliation{Raymond and Beverly Sackler School of Physics and Astronomy, Tel-Aviv University, Tel Aviv, 69978, Israel}

\author{M. Ben Shalom}
\affiliation{Raymond and Beverly Sackler School of Physics and Astronomy, Tel-Aviv University, Tel Aviv, 69978, Israel}

\author{G. Deutscher}
\affiliation{Raymond and Beverly Sackler School of Physics and Astronomy, Tel-Aviv University, Tel Aviv, 69978, Israel}

\author{Y. Dagan}
\affiliation{Raymond and Beverly Sackler School of Physics and Astronomy, Tel-Aviv University, Tel Aviv, 69978, Israel}


\date{\today}

\pacs{}
\keywords{}

\begin{abstract}
 We report transport measurements, including: Hall, Seebeck and Nernst Effect. All these transport properties exhibit anomalous field and temperature dependences, with a change of behavior observed at about $H\sim1.5T$ and $T\sim15K$. We were able to reconcile the low-temperature-low-field behavior of all transport properties using a simple two band analysis. A more detailed model is required in order to explain the high magnetic field regime.
\end{abstract}

\maketitle

\section{INTRODUCTION}
It has been demonstrated that when depositing more than 3 unit cells of \LAO~ over a TiO$_{2}$ terminated \STO~ a conducting layer appears at the interface.\cite{Ohtomo2004, Thiel2006} The nature and origin of the charge carriers are still unknown. Simple electrostatic consideration suggests a transfer of 0.5 electron per unit cell.\cite{Ohtomo2004}Other theoretical work suggests that the electronic reconstruction that occurs at the interface results in lattice deformation and a charge density that corresponds to less than 0.5 electron per unit cell.\cite{pentchevaPicketPRL}
Oxygen vacancies \cite{kalabukhovOVac} and cationic mixing \cite{WillmottPRL} can also contribute to the conductance. The relative contributions of these processes depend on deposition condition. Recent measurements of Shubnikov-de Haas (SdH) oscillations indicate the possibility of a two-band structure.\cite{BenShalom2010} Other works have interpreted the non linear Hall voltage as being the anomalous Hall effect arising from incipient magnetization.\cite{Seri2009, BenShalomPRL}. On the theory side magnetic effects have been predicted to take place at the interfaces.\cite{Zhicheng_Zhong, popovictheoryfortwodeg}
\par
In metals, the thermopower is expected to be linear for T$<$T$_{F}$ and the Nernst signal should, in principle, be zero. This is due to the cancelation of the two terms in the Nernst coefficient: $Q_N=\rho \alpha_{xy}-\rho_{xy}\alpha$, with $\rho$ the resistivity, $\rho_{xy}$ the Hall resistance and $\alpha_{ij}$ the Peltier tensor.\cite{Wang2006} When two types of charge carriers are present this cancelation is relaxed and the Nernst signal can be finite. One therefore expects the Nernst signal to be a sensitive probe to the existence of multiple types of charge carriers. We can therefore use the resistivity, Hall, thermopower and Nernst signal data and see whether the multiple carrier picture can explain the transport. This is what we have done. We find that the low temperature-low field regime can be explained using a simple model of two types of charge carriers. The high field regime for the Nernst measurements cannot be explained by this approach.

\begin{figure}
  \subfigure[]{\label{fig:RT}\includegraphics[width=0.5\textwidth]{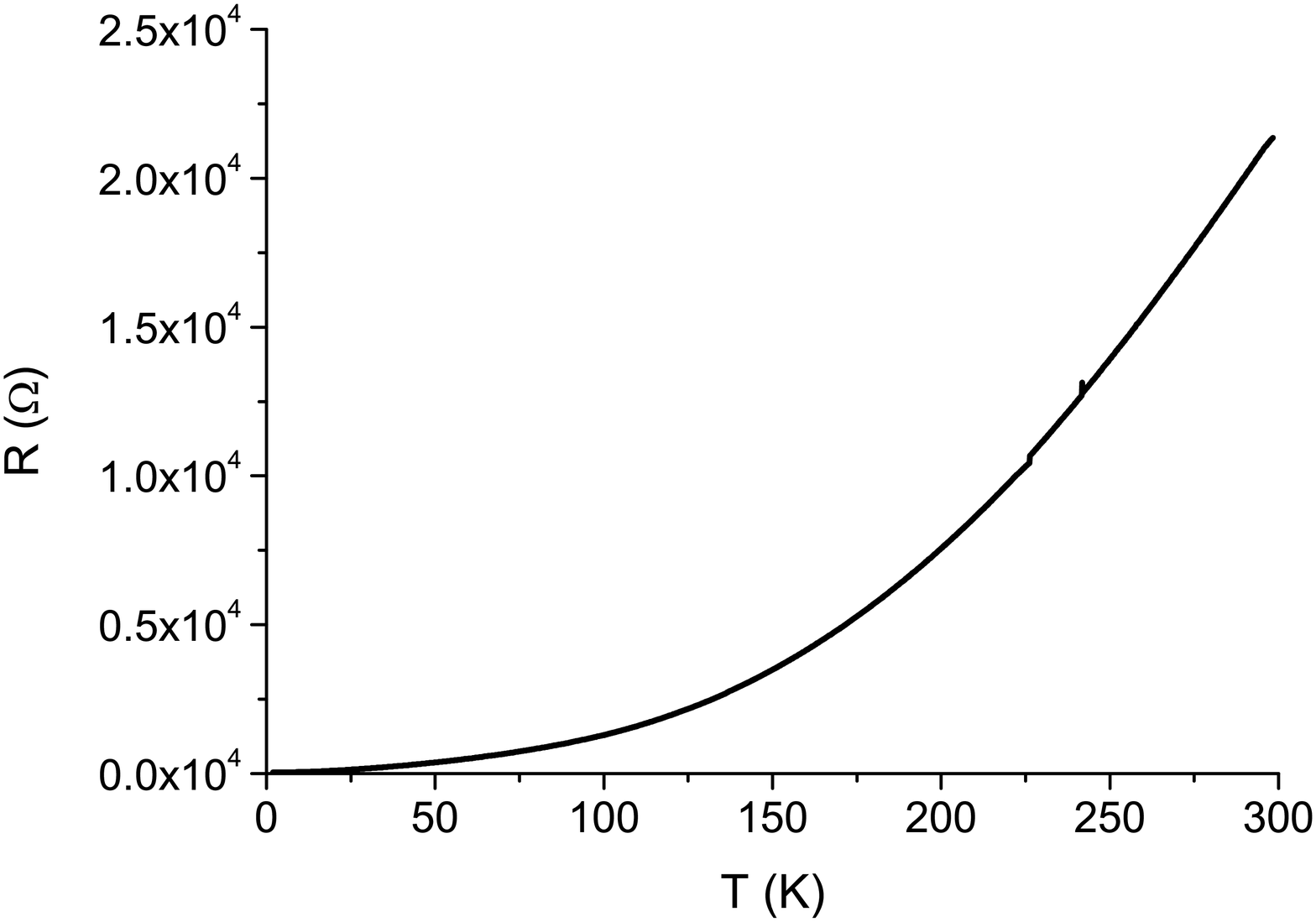}}
  \subfigure[]{\label{fig:MR}\includegraphics[width=0.5\textwidth]{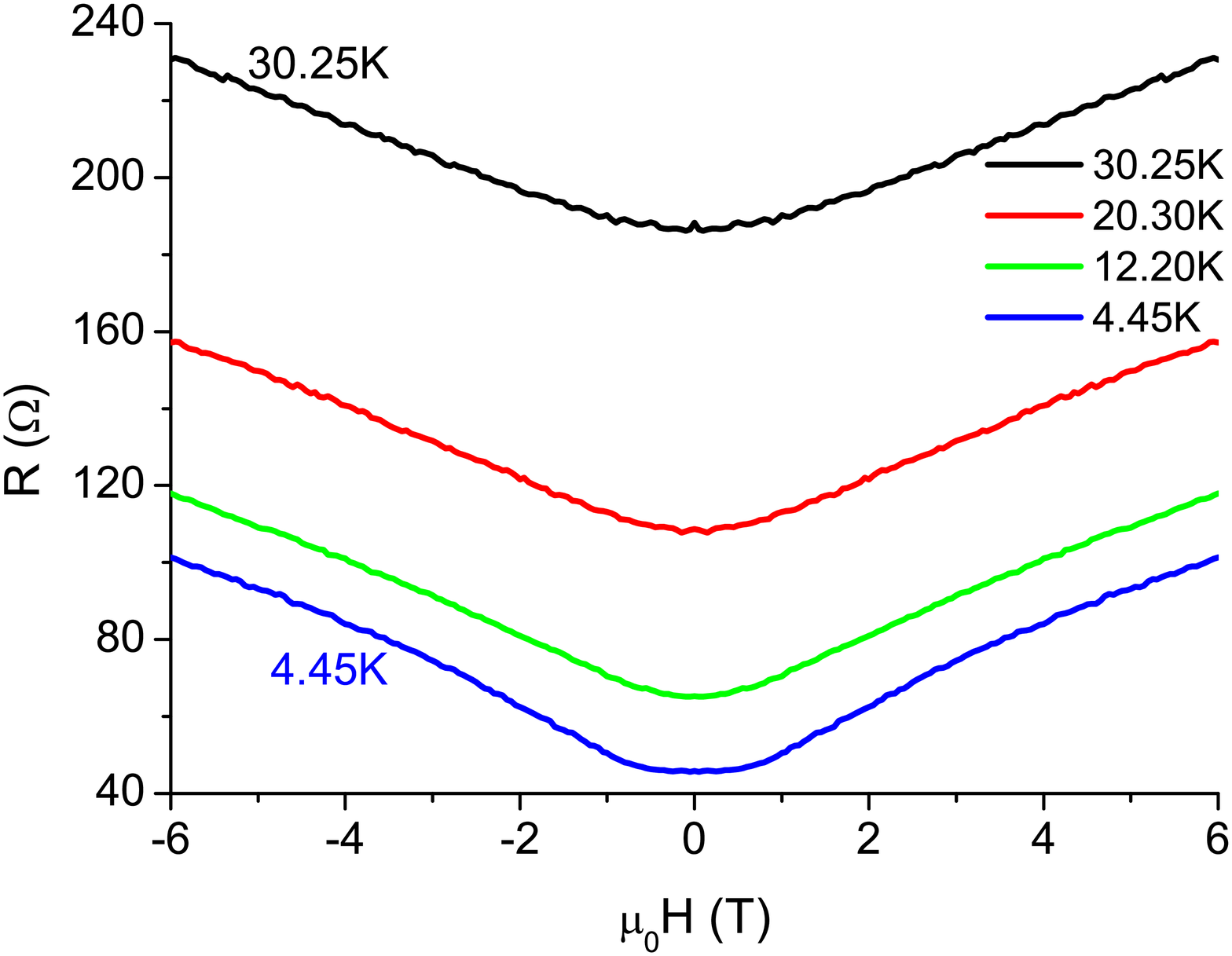}}
  \caption{(colour online) (a) Electrical resistance as a function of temperature. (b) Electrical resistance as a function of applied perpendicular magnetic field.} 
  \label{fig:elecTrans01}
\end{figure}

\par 
\section{SAMPLE AND EXPERIMENTAL SETUP}
We have used a sample with 17 unit cells of \LAO, deposited by pulsed laser deposition on a TiO$_2$-terminated \STO (100) substrate. The deposition conditions are similar to previously reported samples.\cite{BenShalomPRL} The 2DEG's contacts consisted of Al bond wires.
A custom built probe was used in conjunction with a QD PPMS system. The probe consisted of twined phosphor bronze wires, each pair inside a thin stainless steel pipe used for screening. The sample was mounted onto a copper mechanical clamp at one end while the other end was 'floating' (thermally uncoupled) and the sample chamber itself was pumped to $\sim$1Torr to avoid thermalization of the sample in order to maintain the required thermal gradient. Two Lakeshore Cernox thermometers were used for the temperature gradient measurement. A 220$\Omega$ SMT resistor was used as a heater. We used GE-Varnish for attaching the thermometers and the heater onto the sample's surface. A Keithley 6220/6221 delta-mode DMM was used for the 4 point Hall and resistance measurements. A Keithley 1801 nanovolt preamp was used with a 2001 DMM for the thermal measurements. Each transport measurement was followed by a measurement of the Nernst signal at the same base temperature (all measurements were done sequentially).

When measuring the Nernst signal the sample is set to a certain base temperature and the heater is turned on, allowing for the system to reach a steady state, then the magnet is sweeped from -6T to 6T. We made sure that the zero field temperature returned to it's original value after the field sweep. A thermopower measurement of the 2DEG was conducted between 30K and 3K at zero field (The thermopower has a negligible field dependence at low temperatures).

\begin{figure}
  \subfigure[]{\label{fig:Hall}\includegraphics[width=0.5\textwidth]{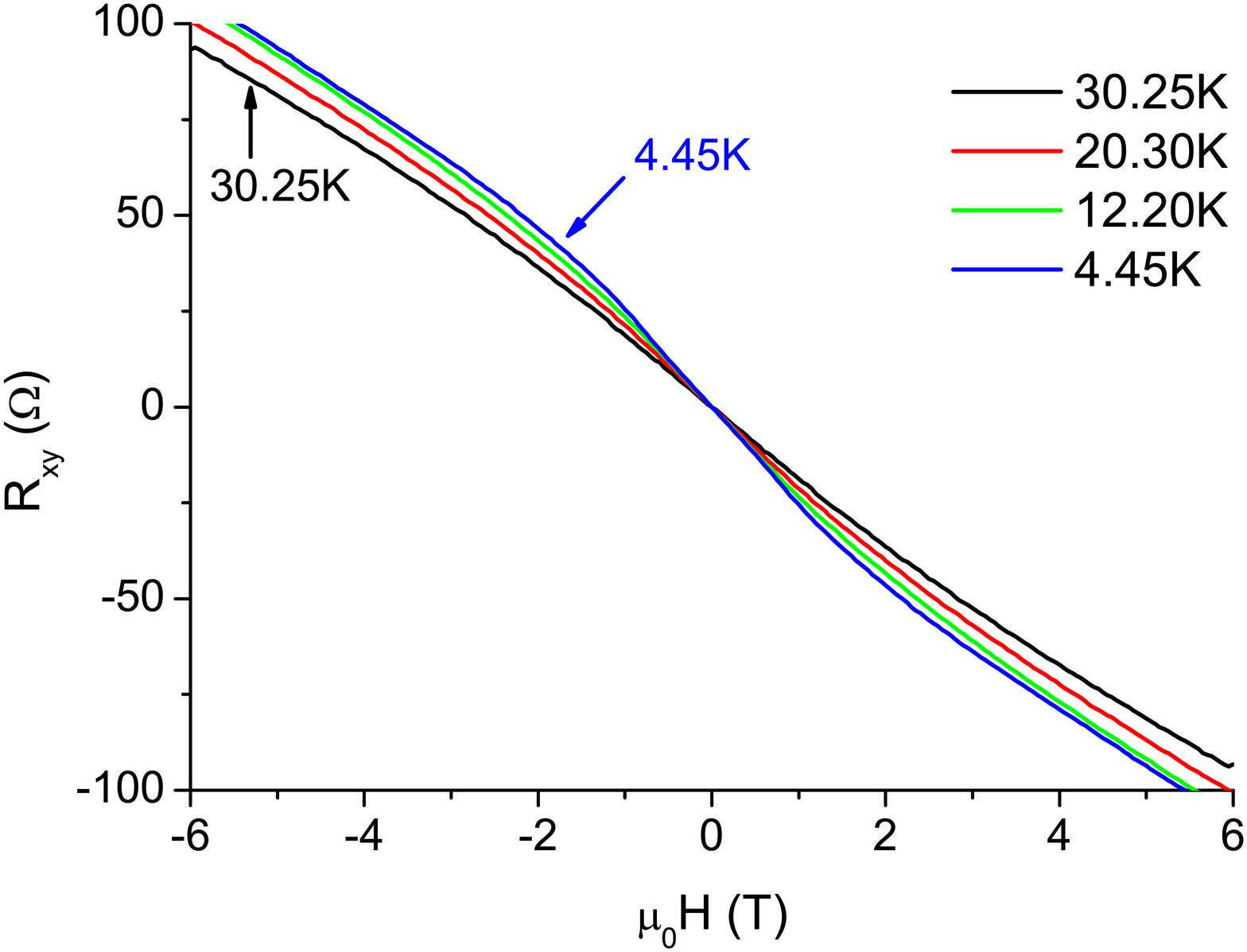}}
  \subfigure[]{\label{fig:HallFit}\includegraphics[width=0.5\textwidth]{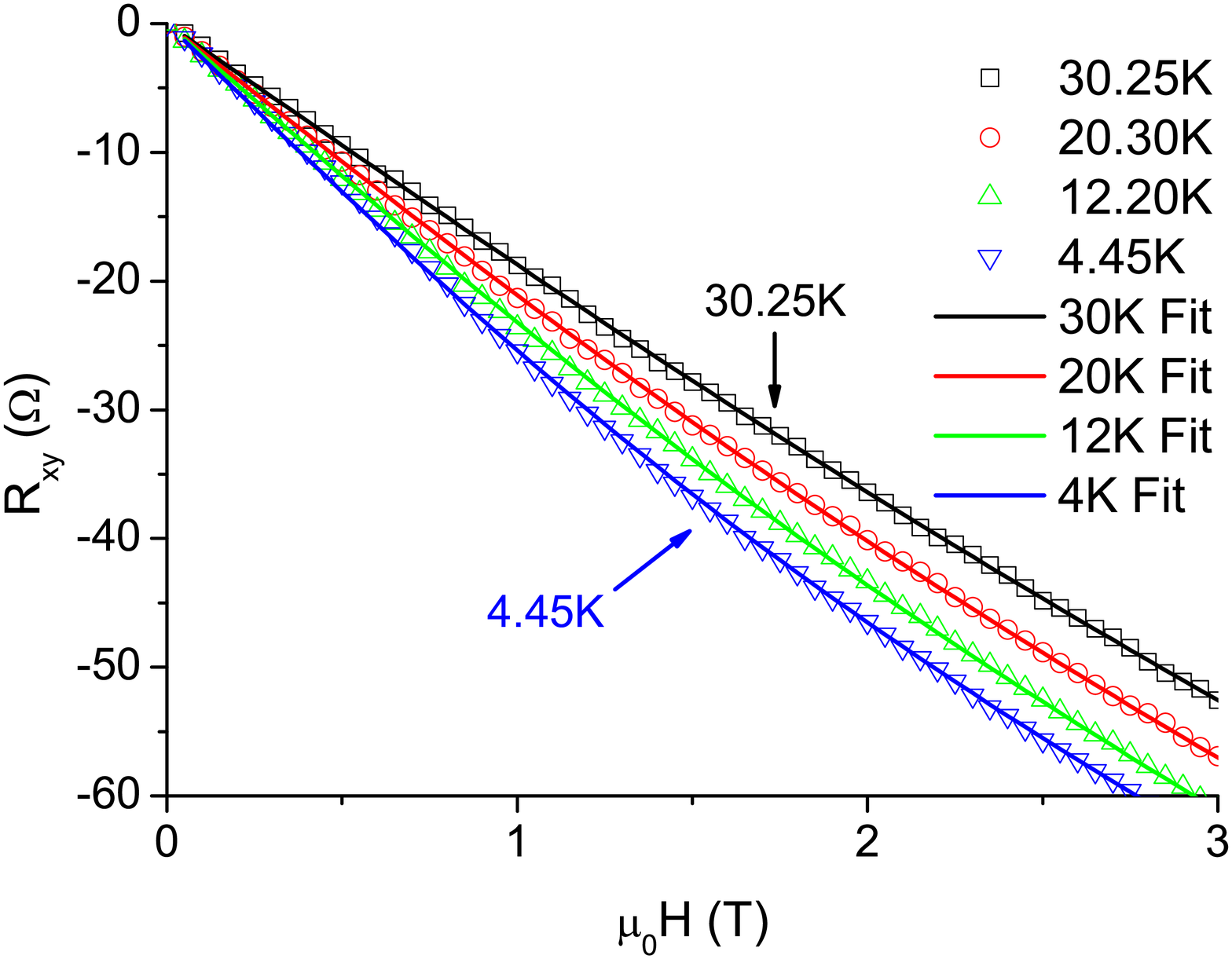}}
  \caption{(colour online) (a) $\rho_{xy}$ as a function of applied perpendicular magnetic field. (b) Fit of the $\rho_{xy}(H)$ measurements using a model of multiple types of charge carriers. The fit was done for the field range of -6T to 6T.}
  \label{fig:elecTrans02}
\end{figure}

\section{DATA AND ANALYSIS}
In Fig \ref{fig:elecTrans01}-\ref{fig:elecTrans02} we show the longitudinal and transverse resistance as a function of temperature and field. These measurements are consistent with previous results obtained by our group.\cite{benshalom} The anomalous change in slope of the magnetoresistance at low temperatures occurs at $H\sim1.5T$, while the Hall resistivity ($\rho_{xy}$) shown in Fig \ref{fig:Hall} exhibits a distinctive non-linear behavior, having a pronounced ``kink'' also at $H\sim1.5T$.

\begin{figure}
  \subfigure[]{\label{fig:NHighT}\includegraphics[width=0.5\textwidth]{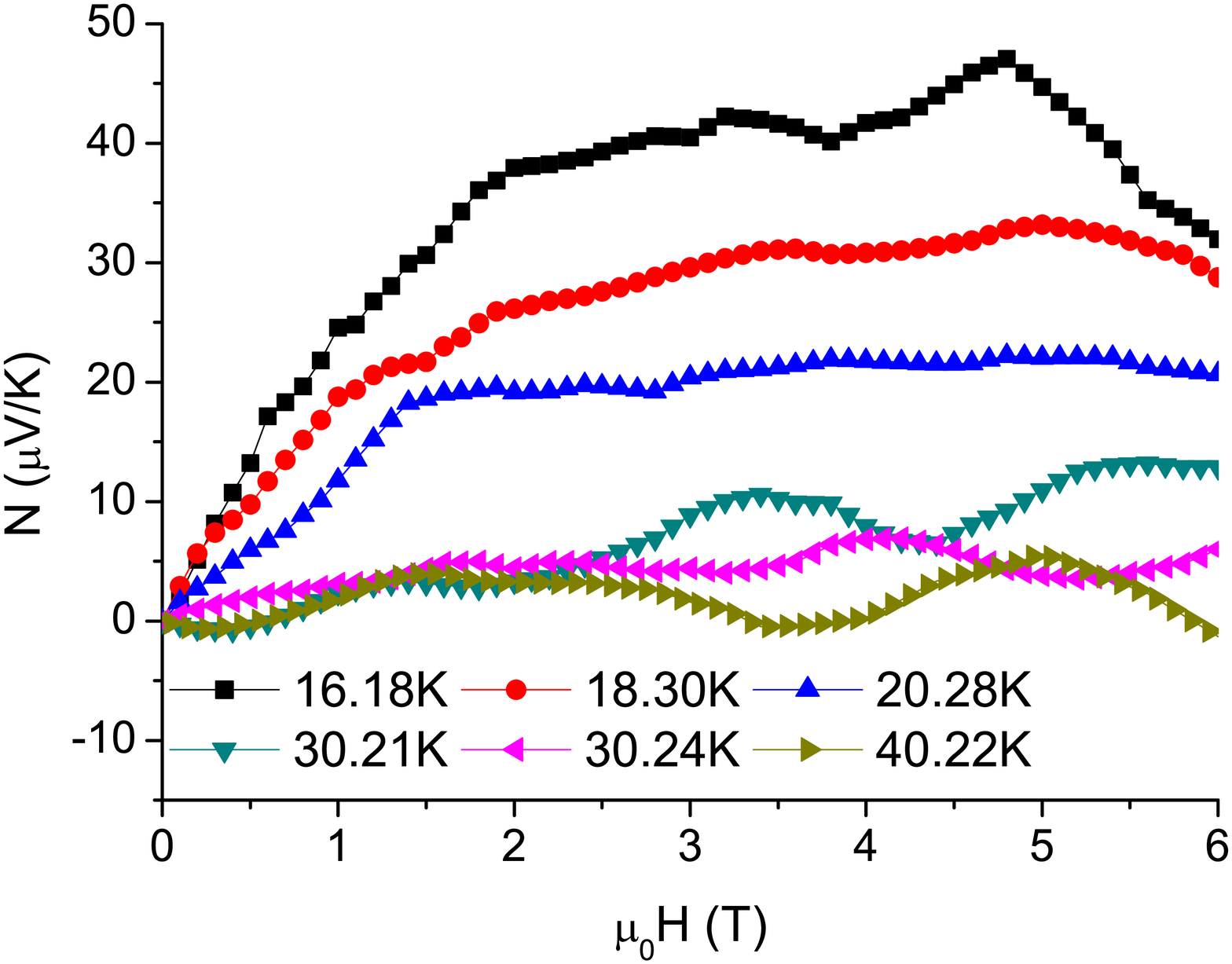}}
  \subfigure[]{\label{fig:NLowT}\includegraphics[width=0.5\textwidth]{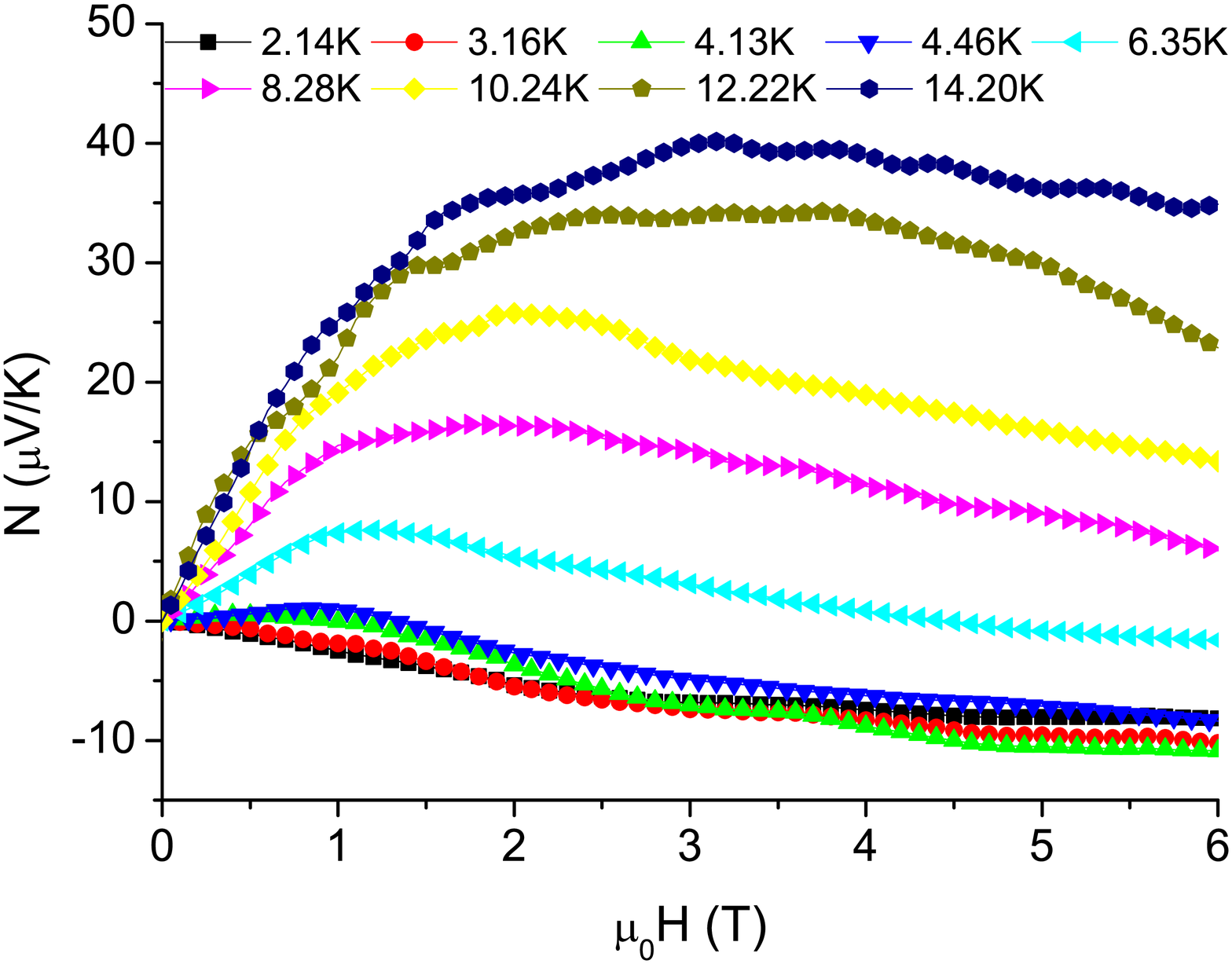}}
  \caption{(colour online) Nernst signal of the \STO-\LAO~ sample;(a) Measurements done for temperature range of $15K<T<40K$. (b) Measurements done for temperature range of $2.2K<T<15K$.}
  \label{fig:thermTrans}
\end{figure}

Fig.\ref{fig:thermTrans} shows the temperature dependence of the Nernst signal taken from the antisymmetric component of the transverse voltage. This eliminates contributions from the contacts and misalignments. As the temperature is lowered the signal increases until a maximum is reached at T=15K. Reducing the temperature further the signal decreases to the point where it changes sign at T=3K for the entire range of field studied, becoming roughly linear.

\par
This anomalous behavior of the Nernst signal is unlikely to be associated with superconductivity or superconducting fluctuations since the critical temperature is $T_c\approx350mK$, while the anomalous signal appears above 6K. In the case of superconducting fluctuations above $T_c$, the Nernst signal is expected to be enhanced significantly when approaching $T_c$.

The next option to explain the large Nernst signal is induced magnetization and the spin-orbit coupling. In this case, due to a mechanism similar to the anomalous Hall effect involving skew scattering, a large Nernst signal is expected\cite{Miyasato2007}. However, for this scenario, one expects the Nernst signal to increase with magnetic field or merely to saturate.

\begin{figure}
  \subfigure[]{\label{fig:NVST}\includegraphics[width=0.5\textwidth]{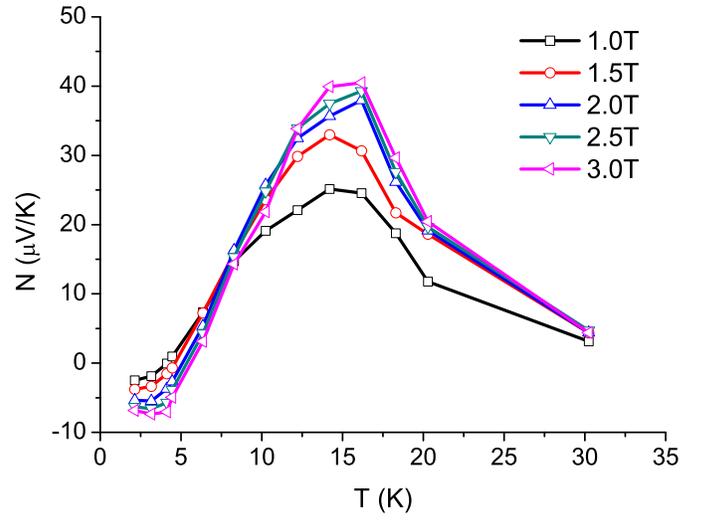}}
  \subfigure[]{\label{fig:TP}\includegraphics[width=0.5\textwidth]{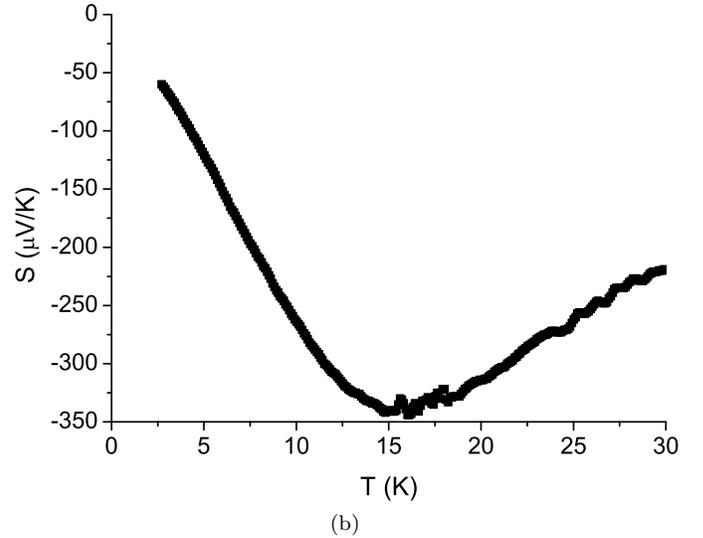}}
  \caption{(colour online) Nernst Signal and thermopower measurements of the \STO-\LAO~ sample;(a) Nernst signal as a function of temperature 30K down to 3K. (b) Measurement of the thermopower as a function of temperature from 30K down to 3K.}
  \label{fig:thermTrans02}
\end{figure}

This expected behavior is not observed in our measurements where the amplitude of the Nernst signal decreases at large fields. In addition, the signal is expected to increase or saturate as the temperature is lowered, in contrast with our measurements in low fields. We therefore conclude that induced magnetization is not the main mechanism responsible for the large Nernst effect for T$>$4K and H$<$2T. One should notice that in contrast to the low-field behavior, the high-field Nernst signal seems to have a temperature independent slope, persisting from 15K down to the lowest temperature measured. This slope dominates the Nernst signal below 3K, while the non monotonic field dependence of the Nernst signal is absent. This will be discussed in the concluding part, from now on we shall be focusing on the low-field regime.

\par
The next option one needs to consider is that the big Nernst signal originates from multiple types of carriers. This possibility had been considered in the past and is in agreement with the non-linear shape of the Hall measurements.\cite{BenShalomPRL} Recent results of Shubnikov de-Haas measurements on \STO-\LAO~ also indicate the existence of two types of carriers.\cite{BenShalom2010} The Nernst signal shows a definite change in slope at a field of $H\sim1.5T$, similarly to magnetoresistance and Hall measurements.

To extract the transport contribution of each band, we analyse the Hall resistivity similarly to Ben-Shalom \etal.\cite{BenShalomPRL} This simplistic model assumes two-types of charge carriers, even though it is possible that a more complicated band structure is involved and other effects might be present. The Hall data is fitted using:

\begin{equation}
\rho_{xy}=\frac{\sigma_1^2R_1+\sigma_2^2R_2+\sigma_1^2\sigma_2^2R_1R_2(R_1+R_2)B^2}{(\sigma_1+\sigma_2)^2+\sigma_1^2\sigma_2^2(R_1+R_2)^2B^2}B,
\label{eq:rhoxy}
\end{equation}

with $R_i$ and $\sigma_i$ are the Hall coefficient and conductivity of the $i^{th}$ type of carrier. Fig \ref{fig:HallFit} shows the fit to $\rho_{xy}$ for the entire field range done using Eq.\ref{eq:rhoxy} and the measured zero field resistance, taken from the magnetoresistance measurements (Fig \ref{fig:MR}). The carrier concentrations and conductance contribution of each band were then extracted and are summarized in Table \ref{tab:calcVals} and Fig \ref{fig:Calculated}. 
The discrepancy between the carrier concentration we obtained and SdH experiments, has already been noted and explained by possible multiple degeneracy scenario.\cite{BenShalom2010}

\begin{figure}
  \begin{center}
    \subfigure[]{\label{fig:Sigma}\includegraphics[width=0.5\textwidth]{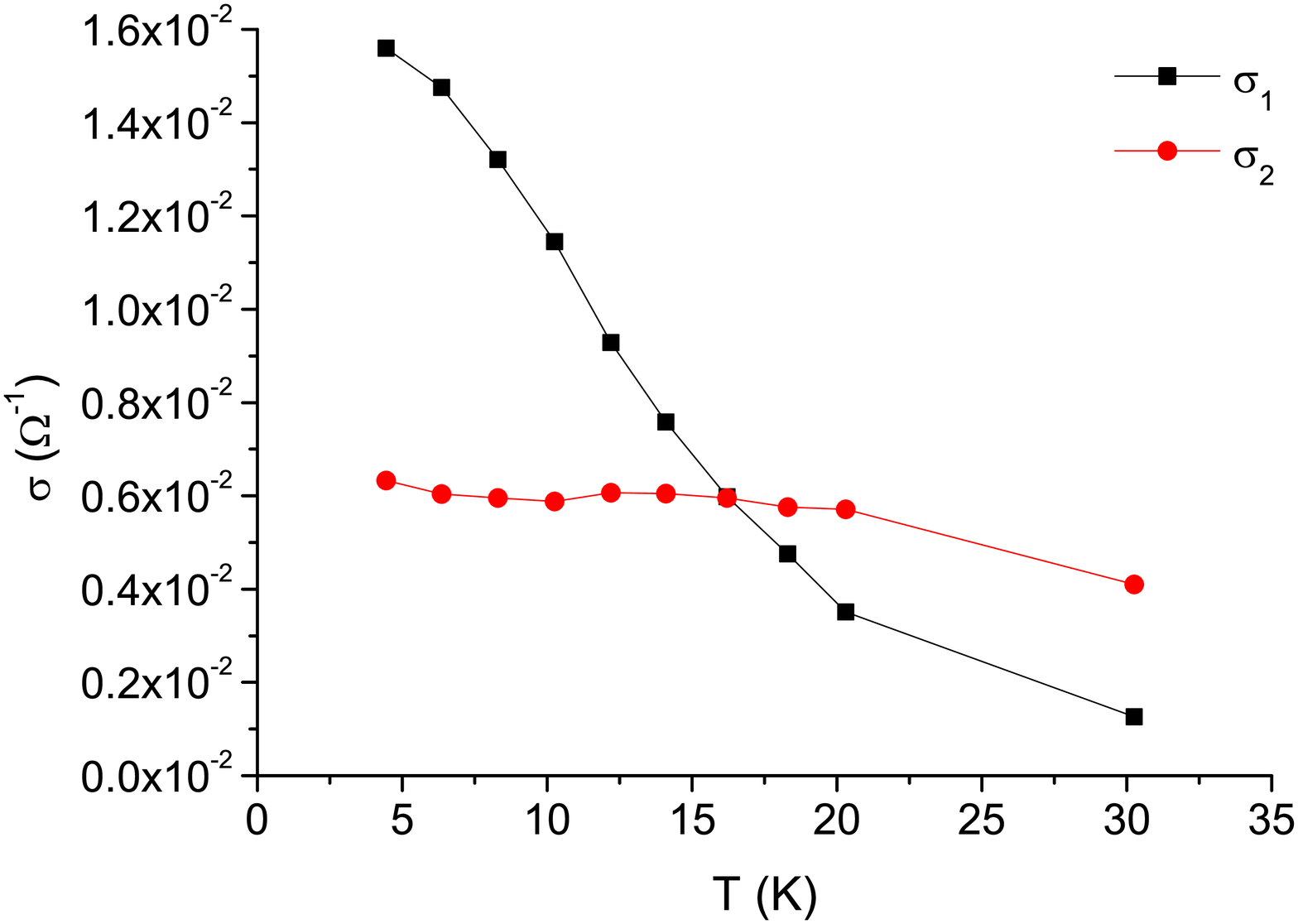}}
    \subfigure[]{\label{fig:RHall}\includegraphics[width=0.5\textwidth]{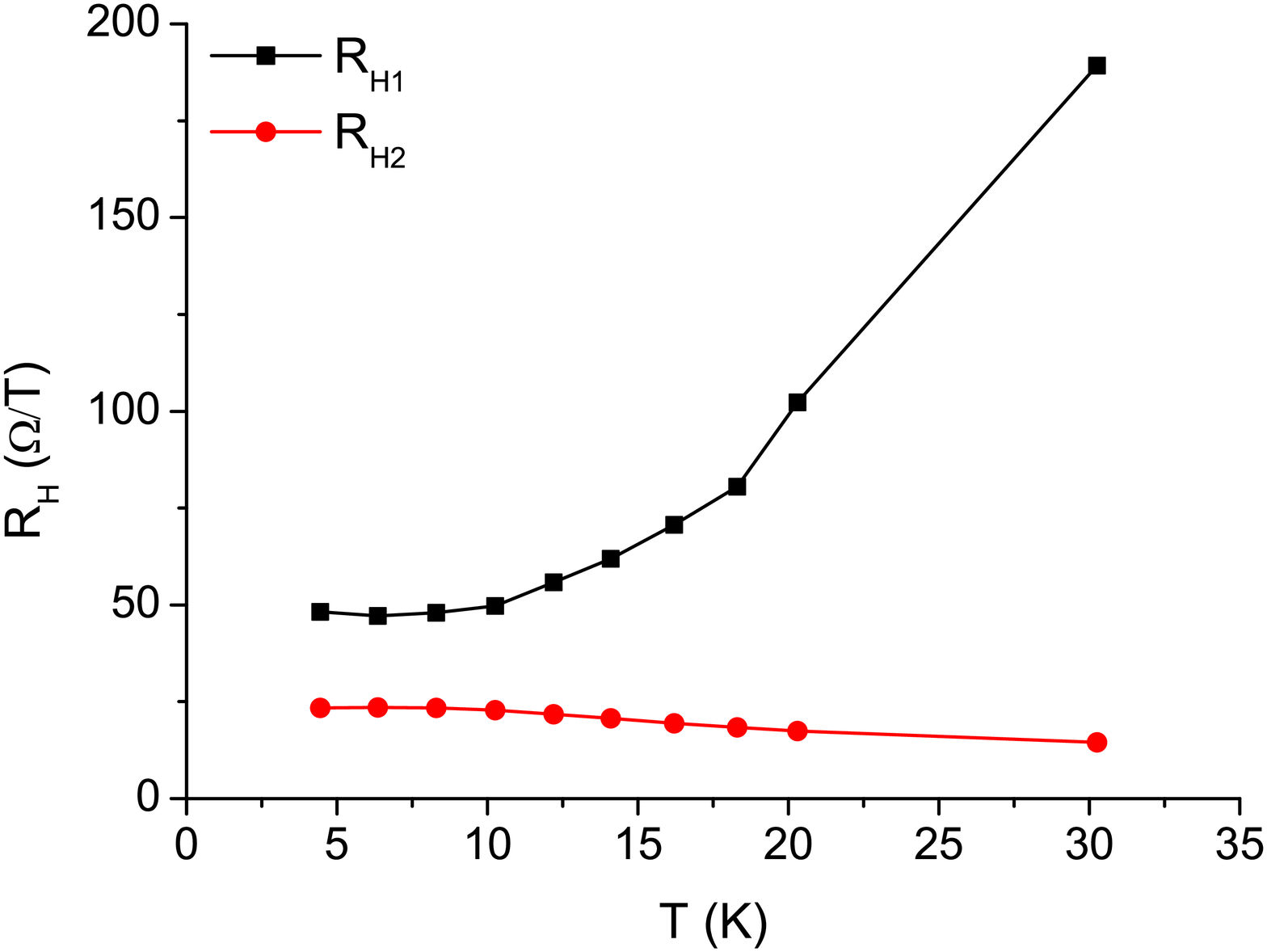}}
  \end{center}
  \caption{(colour online) Calculated electrical properties of the 2DEG assuming a two-band model. (a) Electrical Conduction contributed by each conductance band. (b) Hall coefficent of each conductance band.}
  \label{fig:Calculated}
\end{figure}

\begin{figure}
  \begin{center}
    \subfigure[]{\label{fig:nDensity}\includegraphics[width=0.5\textwidth]{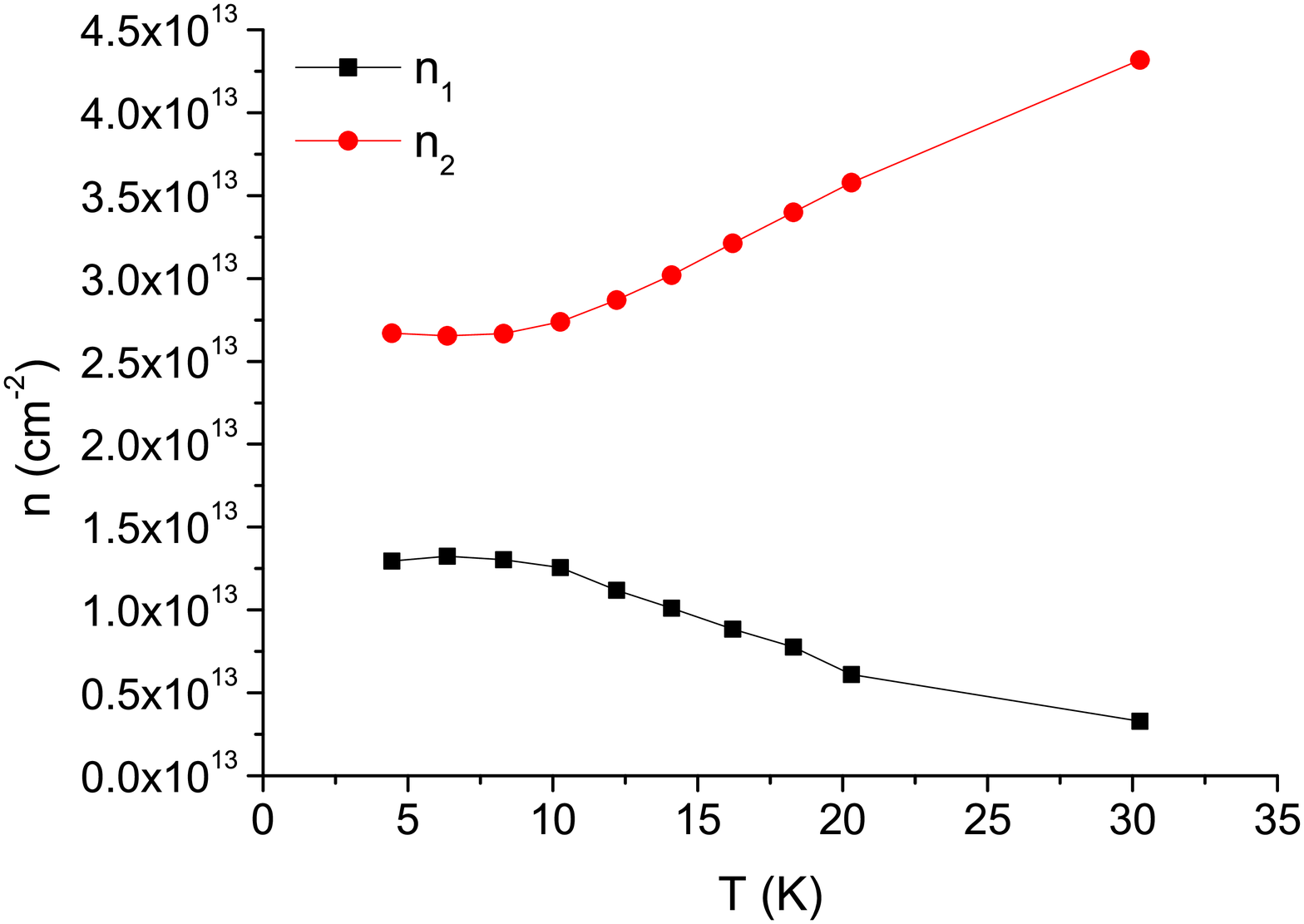}}
    \subfigure[]{\label{fig:NernstZoom}\includegraphics[width=0.5\textwidth]{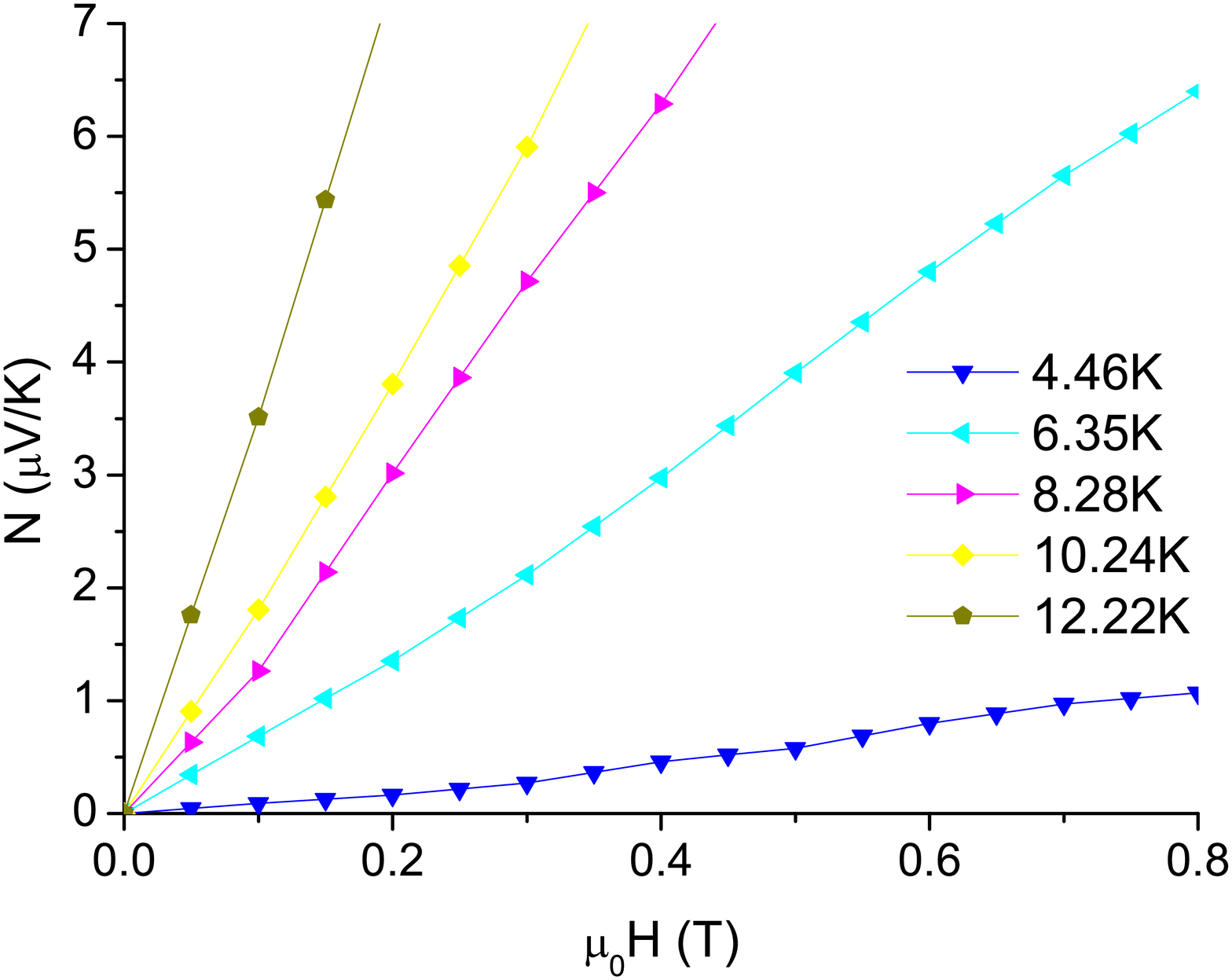}}
  \end{center}
  \caption{(colour online) (a) Density of charge carriers of each conductance band.(b) Linear part of the Nernst signal at low temperatures and low magnetic fields, used for the calculation of the Nernst coefficients ($Q_i$).}
  \label{fig:Calculated}
\end{figure}

\begin{table*}
  \begin{tabular}{llllllllllll}
  \hline
  \hline
  T & [K] & 4.45 & 6.35 & 8.3 & 10.25 & 12.2 & 14.1 & 16.2 & 18.3 & 20.3 & 30.25\\
  \hline
  R$_{H1}$ & [$\Omega $T$^{-1}$] & 48.26 & 47.15 & 47.96 & 49.73 & 55.82 & 61.87 & 70.71 & 80.50 & 102.33 & 189.26\\
  R$_{H2}$ & [$\Omega $T$^{-1}$] & 23.40 & 23.54 & 23.43 & 22.83 & 21.77 & 20.69 & 19.45 & 18.38 & 17.46 & 14.47\\
  $\sigma_{1}$ & [$\Omega^{-1}$m$^{-2}$] & 0.0156 & 0.0147 & 0.0132 & 0.0114 & 0.0092 & 0.0075 & 0.0059 & 0.0047 & 0.0035 & 0.0012\\
  $\sigma_{2}$ & [$\Omega^{-1}$m$^{-2}$] & 0.0063 & 0.0060 & 0.0059 & 0.0058 & 0.0061 & 0.0060 & 0.0059 & 0.0058 & 0.0057 & 0.0041\\
  n$_{1}$ & [cm$^{-2}$x10$^{13}$] & 1.295 & 1.325 & 1.303 & 1.256 & 1.119 & 1.010 & 0.884 & 0.776 & 0.610 & 0.330\\
  n$_{2}$ & [cm$^{-2}$x10$^{13}$] & 2.670 & 2.655 & 2.667 & 2.737 & 2.870 & 3.019 & 3.213 & 3.399 & 3.578 & 4.318\\
  \hline
  \hline
  \end{tabular}
  \caption{(colour online) Calculated values of the Hall coefficients, electrical conductivity and density of charge carriers per band, extracted from the non-linear Hall measurement.}
  \label{tab:calcVals}
\end{table*}

\par
\begin{figure}
  \centering
  \subfigure[]{\label{fig:Sfit}\includegraphics[width=0.5\textwidth]{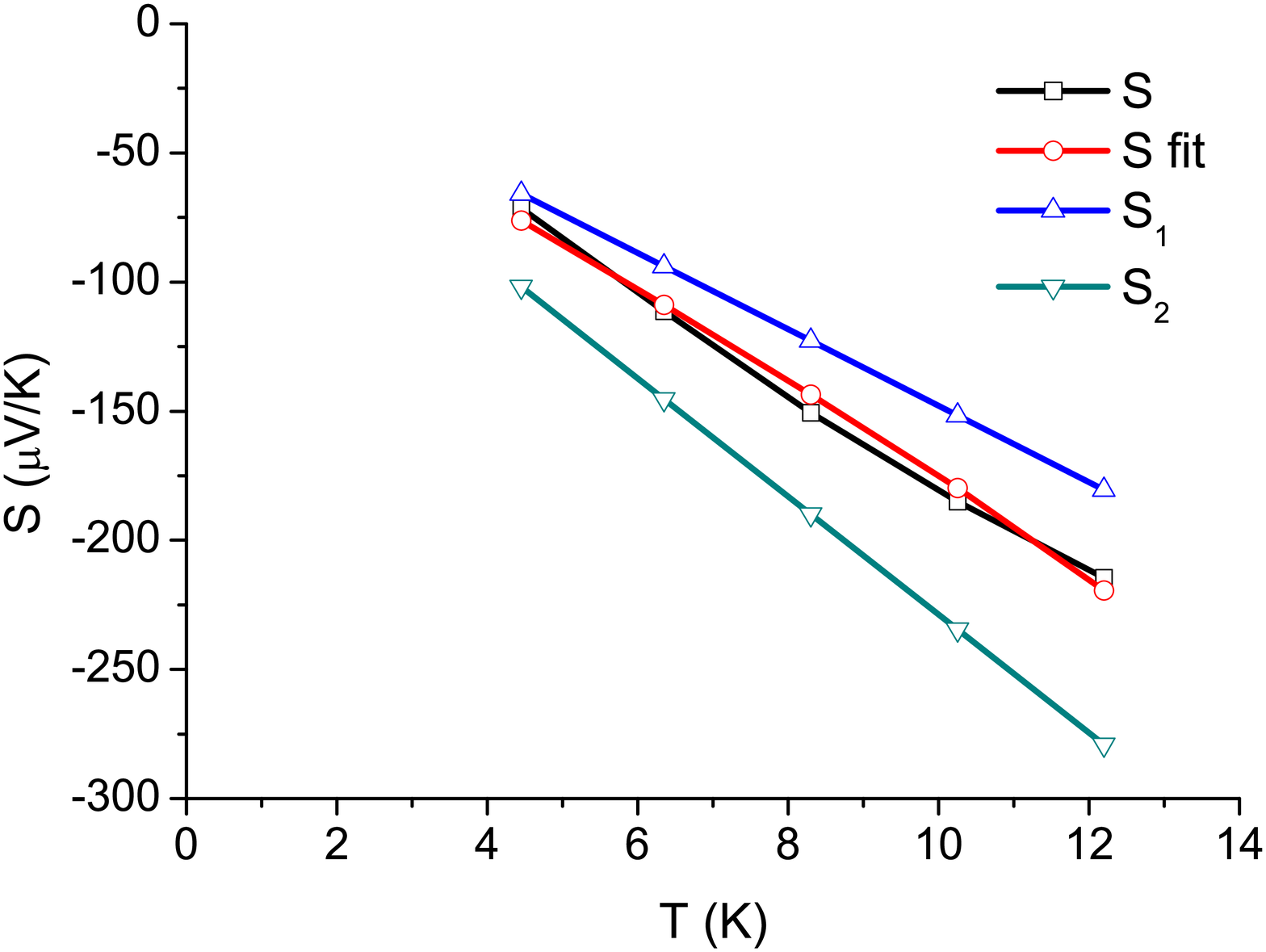}}
  \subfigure[]{\label{fig:Qfit}\includegraphics[width=0.5\textwidth]{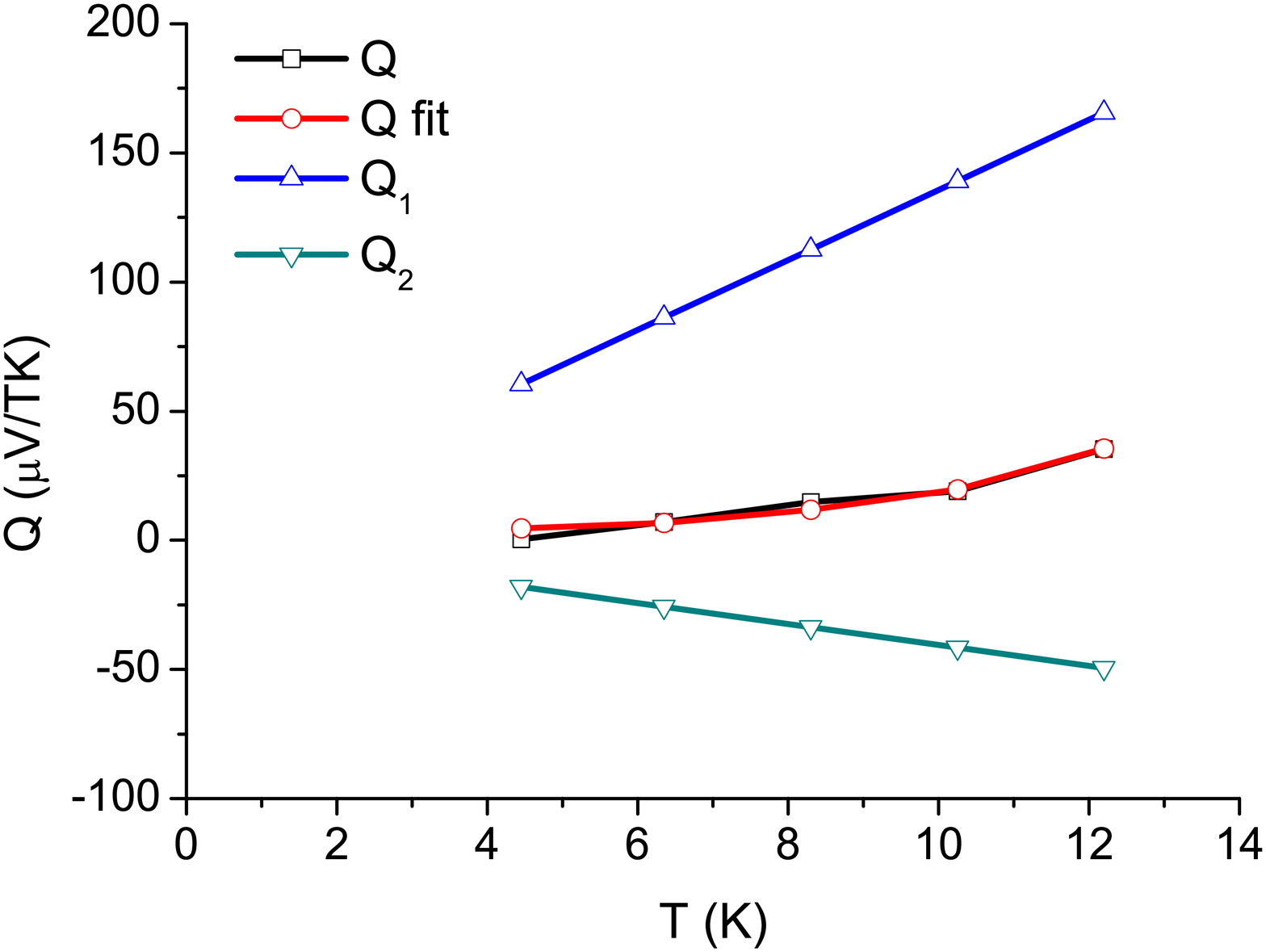}}
  \caption{(colour online) Fit for the thermopower and Nernst coefficient. (a) Black solid line shows the measured thermopower data. Red Solid line shows the calculated fit, while the blue and green lines shows the values associated to each type of charge carriers. (b) Black solid line shows the measured Nernst data. Red solid line shows the calculated fit, while the blue and green lines shows the values associated to each type of charge carriers.}
  \label{fig:fitting}
\end{figure}

 The thermopower measurement is shown in Fig.\ref{fig:TP}. A change in slope is evident at T=15K below which the thermopower turns linear as expected for low temperatures ($T\ll T_F$), and extrapolates linearly to zero at T=0K. In this low-temp regime, we use the simple expression for the case of a degenerate 2D free electron gas:\cite{KARAVOLAS1991}
\begin{equation}
S(T)=-\frac{\pi^2}{3}\frac{k_B}{e}\frac{T}{T_F}(1+\alpha).
\label{eq:T_F}
\end{equation}
We estimate $T_F\sim15K$, assuming the simple case of an energy independent relaxation time ($\alpha=0$), which is in general agreement with the linear temperature dependence observed below 15K. We assume a linear contribution by each type of charge carrier to the thermopower and the Nernst coefficient:

\begin{equation}
S(T)=\frac{S_1 \sigma_1 + S_2 \sigma_2}{(\sigma_1 + \sigma_2)},
\label{eq:TP}
\end{equation}
\begin{equation}
Q(T)=\frac{Q_1 \sigma_1 + Q_2 \sigma_2}{\sigma_1 + \sigma_2} + \frac{\sigma_1 \sigma_2 (S_1 - S_2)(\sigma_1 R_1 - \sigma_2 R_2)}{(\sigma_1 + \sigma_2)^2},
\label{eq:Q}
\end{equation}

with $S_i$ is the thermopower and $Q_i$ the Nernst coefficient conribution of the $i^{th}$ type of carrier.
Using the results extracted from the fits to the Hall data and Eq.\ref{eq:TP}, the parameters of $S_1$ and $S_2$ are then extracted using the $\chi^2$ method\cite{Press2007} and used in Eq.\ref{eq:Q} with the linear slope of the Nernst signal at low fields, to calculate $Q_1$ and $Q_2$. The fit was done assuming a simple model of linear response in field and temperature for both components of the thermopower and the Nernst coefficient. The resulting values for $S_i$ and $Q_i$ are presented in Table \ref{tab:calcTherm}. The fits used for the thermopower and Nernst calculations are shown in Fig \ref{fig:Sfit} and \ref{fig:Qfit} respectively. Eq.\ref{eq:Q} applies to the isothermal Nernst coefficient. In our case, the sample is considered isotropic in the 2DEG plane, and the electron mobility is quite low, so a Righi-Leduc effect should be negligible. Therefore, we can use it for our measured (adiabatic) Nernst coefficient.\cite{Sugihara1969, Wang2001}

\begin{table}
\begin{centering}
  \begin{tabular}{cccc}
  \\
  \hline
  \hline
  $S_1/T~[\frac{\mu V}{K^2}]$ & $S_2/T~[\frac{\mu V}{K^2}]$ & $Q_1/T~[\frac{\mu V}{K^2T}]$ & $Q_2/T~[\frac{\mu V}{K^2T}]$\\
  \hline
  -14.79 & -22.88 & -4.06 & 13.57\\
  \hline
  \hline
  \end{tabular}
  \caption{(colour online) Calculated values of the thermopower (S$_i$) and Nernst coefficient (Q$_i$) for each band.}
  \label{tab:calcTherm}
\end{centering}
\end{table}

The conductance, Hall coefficients and charge carrier densities, calculated from the electrical transport data indicate a change in behavior at $T\approx15K$ in agreement with the thermal transport measurements (Fig \ref{fig:thermTrans}).
As seen in Fig \ref{fig:nDensity},\ref{fig:Sigma}, the density of the two types of charge carriers remains roughly unchanged, while their contribution to the conductance is not, which implies that the change in conductance is related to the carriers' mobility.\cite{Bell2009} More so, Fig \ref{fig:Sigma} shows that one type of charge carriers' contribution to the 2DEG's conductance is relatively temperature independent, while the other type of carrier varies significantly.

\section{DISCUSSION AND SUMMARY}
Our results indicate the magnetic field scale of $H\sim1.5T$ and the temperature of $T\sim15K$ as points where the system undergoes a distinct change in behavior, perhaps due to a change in the contribution of one of the two bands of charge carriers. This behavior is evident in the thermopower, where a minima is observed at 15K. Below this temperature the thermopower becomes linear. At this temperature range we can see that the Nernst signal at high magnetic field turns linear. As the temperature is lowered, the linear regime broadens, spanning the entire field range below 3K. This linear Nernst signal is anomalously large. It may arise from dominance of the mobile band, as indicated from the conductance shown in Fig.\ref{fig:Sigma}. Another possibility is that a magnetic order is induced at high fields and low temperatures, enhancing the Nernst signal. In this view, the minima in the thermopower maybe related to a Kondo-like temperature. However, in a simple Kondo scenario one would expect a minimum in resistivity which is absent in our data.
It is clear that the Nernst signal is unrelated to superconducting fluctuations. Further experimental study is needed in order to determine which carriers undergo a superconducting condensation and what is the role of fluctuations at lower temperatures. We speculate that the more mobile band, which is observed in SdH measurements,\cite{BenShalom2010, Caviglia2010} is responsible for the superconducting transition as we can see that its Nernst signal becomes dominant at low temperatures (below 5K). 

\par In summary, transport measurements of the 2DEG at the interface of a \STO-\LAO~ show an anomalous behavior which cannot be explained by a single type of charge carriers. The electrical transport properties are consistent with previous results, specifically the carrier concentrations and the electron mobility, obtained by Hall effect.\cite{BenShalomPRL, BenShalom2010} We were able to successfully fit the measured Hall resistivity as well as the thermopower at low temperature and the corresponding Nernst coefficient in low magnetic fields to a two-band system model. The electrical and thermal measurements are in excellent agreement.
\par
\section{ACKNOWLEDGMENTS}
We thank R. Mor, A. Palevsky and Y. Korenblit for helpful discussions.
This research was supported by the ISF under Grants No. 1421/08, 1543/08.
\bibliographystyle{apsrev}
\bibliography{Paper01}
\end{document}